\documentclass[aps,showpacs]{revtex4}
\usepackage{graphics}
\usepackage{graphicx}

\def\ss{\scriptscriptstyle }

\begin{document}
\title{Gain without inversion in a biased superlattice}

\author{F.T. Vasko$^*$}
\affiliation{NMRC, University College Cork, Lee Maltings
\\
Prospect Row, Cork, Ireland} 
\date{\today}

\begin{abstract}
Intersubband transitions in a superlattice under homogeneous electric field
is studied within the tight-binding approximation. Since the levels are 
equi-populated, the non-zero response appears beyond the Born approximation. 
Calculations are performed in the resonant approximation with scattering 
processes exactly taken into account. The absorption coefficient is equal zero 
for the resonant excitation while a negative absorption (gain without 
inversion) takes place below the resonance. A detectable gain in the THz 
spectral region is obtained for the low-doped $GaAs$-based superlattice and 
spectral dependencies are analyzed taking into account the interplay 
between homogeneous and inhomogeneous mechanisms of broadening.
\end{abstract}

\maketitle

The examination of stimulated emission due to intersubband transitions of 
electrons (monopolar laser effect), which has been carried out during the 
previous decade, have resulted in mid-IR lasers (see Refs. in \cite{1,2}). 
Recently, the THz laser has also been demonstrated \cite{3,4,5,6}. The standard laser 
scheme based on vertical transport through the quantum cascade structures, 
which incorporates the injector and active regions, has been used in 
both cases. Population inversion appears in the active regions and leads 
to stimulated emission for the mode propagating along mid-IR or 
THz waveguide. In contrast to this, the vertical current in a biased 
superlattice (BSL) with the Wannie-Stark ladder, which appears under the 
condition $2T\ll\varepsilon_{\ss B}$ \cite{7} (here $\varepsilon_{\ss B}/\hbar$
is the Bloch frequency and $T$ stands for the tunneling matrix element between 
adjacent QWs), does not change the populations of the levels. Due to this, the 
consideration based on the golden rule approach gives a zero absorption. At 
the same time, for the wide minigap SL, with the width $2T\gg
\varepsilon_{\ss B}$, a negative differential conductivity, i.e. gain due 
to Bloch oscillations, takes place \cite{8}. This contradiction and the
question about THz gain without inversion are discussed in Ref. \cite{9}. 
Last year, agreement between the numerical results for the wide-miniband 
and hopping regimes of high-frequency response was noted in \cite{10}.

Since there is no well-defined dispersion relation between energy and
momentum, $\varepsilon$ and ${\bf p}$, beyond the Born approximation, one
has to consider the intersubband transitions based on the spectral density
function, $A_{\varepsilon}({\bf p})$, which is a finite-width peak \cite{11}. 
Let us consider first the two-level model with an identical distribution 
finction for both levels, $f_{\varepsilon}$. We take into account the
off-resonant transitions with a non-zero detuning energy $\Delta\varepsilon 
=\hbar\omega -\varepsilon_{\ss B}$ with respect to the level splitting energy, 
$\varepsilon_{\ss B}$, see Fig.1. The intersubband absorption is given by
\begin{equation}
\label{1} 
\alpha_{\Delta\varepsilon}\propto\int\frac{d{\bf p}}{(2\pi\hbar )^2}
\int_{-\infty}^{\infty}d\varepsilon A_{\varepsilon}({\bf p})A_{\varepsilon -
\Delta\varepsilon}({\bf p}) (f_{\varepsilon -\Delta\varepsilon}-
f_{\varepsilon}) ,
\end{equation}
moreover the relation $\alpha_{\Delta\varepsilon}\propto{\rm sign}(\Delta
\varepsilon )$ is obtained under the replacement $\varepsilon -\Delta
\varepsilon\rightarrow\varepsilon$. Taking into account that $f_{\varepsilon}$ 
decreases with $\varepsilon$, one immediately obtains $\alpha_{\Delta
\varepsilon}<0$ if $\Delta\varepsilon <0$ \cite{12}, i.e. a gain appears in 
the BSL with disorder beyond the Born approximation. In the Born approximation,
when the spectral function is replaced by $\delta$-function, one obtains 
$\alpha_{\Delta\varepsilon}=0$.

In this paper, we evaluate Eq.(1) for the low-doped BSL taking into account 
the scattering processes exactly. Numerical estimates are performed below 
taking into account the interplay between homogeneous and inhomogeneous 
mechanisms of broadening.
\begin{figure}
\begin{center}
\includegraphics{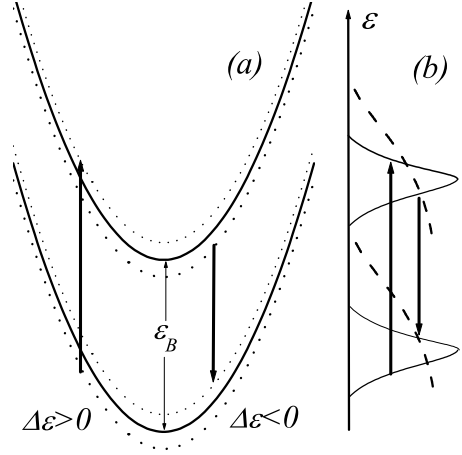}
\end{center}
\addvspace{-1 cm}\caption{Off-resonant intersubband transitions $(a)$ and 
corresponding spectral density functions $(b)$. The dashed curves show 
distribution functions and arrows indicate the transitions with positive
and negative detuning energies.}
\label{fig.1}
\end{figure}

Within the framework of the tight-binding approach we describe the electron 
states in BSL using the matrix Hamiltonian:
\begin{equation}
\label{2} 
\hat{h}_{rr'}=\left(\frac{\hat{p}^2}{2m}+V_{r{\bf x}}+r
\varepsilon_{\ss B}\right) \delta_{rr'}+T(\delta_{rr'-1}+\delta_{rr'+1}) ,
\end{equation}	
where $\hat{p}^2/2m$ is the in-plane kinetic energy operator, $m$ is the 
effective mass, $V_{r{\bf x}}$ is a random potential energy in the $r$-th QW,
$r=0,\pm 1,\ldots$, which are statistically independent in each QW. The Bloch 
energy, $\varepsilon_{\ss B}\simeq |e|FZ$, appears in (2) due to the shift of 
levels in the SL with period $Z$ under a homogeneous electric field $F$, see 
\cite{13}. The perturbation operator due to a high-frequency transverse field 
$[E_{\ss\bot}\exp (-i\omega t)+c.c.]$ is written as $[\widehat{\delta h}_{rr'}
\exp (-i\omega t)+H.c.]$, where the non-diagonal matrix $\widehat{\delta h}_{r
r'}$ is given by
\begin{equation}
\label{3} 
\widehat{\delta h}_{rr'}=\frac{ev_{\ss\bot}}{\omega}E_{\ss\bot}
(\delta_{rr'-1}+\delta_{rr'+1})
\end{equation}	
and $v_{\ss\bot}=TZ/\hbar$. The high-frequency current induced by the 
perturbation (3), $[I_{\omega}\exp (-i\omega t)+c.c.]$, is determined by the 
standard formula:
\begin{equation}
\label{4} 
I_{\omega}=i\frac{2ev_{\ss\bot}}{L^3}\left\langle\left\langle\sum_r
{\rm sp}_{\ss\|}(\widehat{\delta\rho}_{r+1r}-\widehat{\delta\rho}_{r-1r})
\right\rangle\right\rangle ,
\end{equation}	
where 2 is due to spin, ${\rm sp}_{\ss\|}\ldots$ is the averaging over in-plane
motion, $\langle\langle\ldots\rangle\rangle$ is the averaging over random 
potentials $V_{r{\bf x}}$, and $L^3$ is the normalization volume. 

The high-frequency contribution to the density matrix in Eq.(4), $[\widehat
{\delta\rho}_{rr'}\exp (-i\omega t)+H.c.]$, is governed by 
the linearized equation:
\begin{eqnarray}
\label{5} 
-i\omega\widehat{\delta\rho}_{rr'}+\frac{i}{\hbar}(\hat{h}_r\widehat{\delta
\rho}_{rr'}-\widehat{\delta\rho}_{rr'}\hat{h}_{r'}) \nonumber  \\
+\frac{T}{\hbar}(\widehat{\delta\rho}_{r+1r'}+\widehat{\delta\rho}_{r-1r'}
-\widehat{\delta\rho}_{rr'-1}-\widehat{\delta\rho}_{rr'+1}) \nonumber \\
+\frac{i}{\hbar}\widehat{\delta h}_{rr'}(\hat{\rho}_{r'}-\hat{\rho}_{r})=0 . 
\end{eqnarray}	
Here $\hat{h}_r=\hat{p}^2/2m+V_{r{\bf x}}+r\varepsilon_{\ss B}$ describes 
an in-plane motion in the $r$-th QW and we use the steady state density matrix
$(\hat{\rho}_o)_{rr'}\simeq\delta_{rr'}\hat{\rho}_{r}$ , i.e. we have 
neglected a weak non-diagonal term which is responsible for the tunneling
current through the BSL. We restrict ourselves to the consideration of $\propto 
T^2$ contributions only, so that we can omit $\propto T$ addendums in Eq.(5). 
Thus, an independent equation for $\widehat{\delta\rho}_r^{\ss (\pm )}\equiv
\widehat{\delta\rho}_{r\pm 1r}$ takes the form:
\begin{eqnarray}
\label{6} 
-i\omega\widehat{\delta\rho}_r^{\ss (\pm )}+\frac{i}{\hbar}(\hat{h}_{r\pm 1}
\widehat{\delta\rho}_r^{\ss (\pm )}-\widehat{\delta\rho}_r^{\ss (\pm )}
\hat{h}_r) \nonumber \\
\simeq -i\frac{ev_{\ss\bot}}{\hbar\omega}E_{\ss\bot}(\hat{\rho}_{r\pm 1}
-\hat{\rho}_r) .
\end{eqnarray}	
Note that for the collisionless case $(\hat{\rho}_{r\pm 1}-\hat{\rho}_r)
\rightarrow 0$, so that the response vanishes and $I_{\omega}$ is only
non-zero due to differences in scattering processes for adjacent QWs. In 
addition, for the resonant approximation, $|\hbar\omega -\varepsilon_{\ss B} |
\ll\varepsilon_{\ss B}$, one can neglect the contribution of $\widehat{\delta
\rho}_r^{\ss (-)}$.

Writing ${\rm sp}_{\ss\|}\ldots$ in Eq.(4) in the coordinate representation,
we obtain $I_{\omega}=(i2ev_{\ss\bot}/L^3)\langle\sum_r\int d{\bf x}
{\delta\rho}_r^{\ss (+)}({\bf x},{\bf x})\rangle$. Next, we describe the 
electron states in the $r$-th QW by the use of the eigenstate problem
$(\hat{p}^2/2m+V_{r{\bf x}})\psi_{r{\bf x}}^{\nu}=\varepsilon_{r\nu}\psi_{r
{\bf x}}^{\nu}$, where a quantum number $\nu$ marks an in-plane state. Using 
this basis, we transform $I_{\omega}$ into
\begin{equation}
\label{7} 
I_{\omega}\simeq i\frac{2ev_{\ss\bot}}{L^3}\left\langle\left\langle
\sum_{r\nu\nu'}\delta\rho_r^{\ss (+)}(\nu ,\nu')\int d{\bf x}\psi_{r+1
{\bf x}}^{r\nu~*}\psi_{r{\bf x}}^{\nu'}\right\rangle\right\rangle 
\end{equation}	
and the linearized kinetic equation takes the form:
\begin{eqnarray}
\label{8} 
(\varepsilon_{r+1\nu}-\varepsilon_{r\nu'}+\varepsilon_{\ss B}-\hbar\omega-i
\lambda )\delta\rho_r^{\ss (+)}(\nu ,\nu') \nonumber \\
=\frac{ev_{\ss\bot}}{\hbar\omega}E_{\ss\bot}[f_{\varepsilon_{r+1\nu}}-
f_{\varepsilon_{r\nu'}}]\int d{\bf x}\psi_{r+1{\bf x}}^{\nu}\psi_{r{\bf x}}
^{\nu'~*} . 
\end{eqnarray}
Here $\lambda\rightarrow +0$ and we use the quasi-equilibrium distribution 
$\hat{\rho}_r=f_{\hat{p}^2/2m+V_{r{\bf x}}}$, where $f_{\varepsilon }$ is the 
Fermi function with identical chemical potentials, $\mu$, and temperatures,
$T_e$, for any QW. We introduce the conductivity, $\sigma_{\omega}$, according 
to $I_{\omega}=\sigma_{\omega}E_{\ss\bot}$, and Eqs. (7,8) give us:
\begin{equation}
\label{9} 
\sigma_{\omega}=i\frac{2(ev_{\ss\bot})^2}{\omega L^3}\left\langle
\left\langle\sum_{r\nu\nu'}\frac{(f_{\varepsilon_{r+1\nu}}-f_{\varepsilon_{r
\nu'}})Q_{r+1,r}^{\nu\nu '}}{\varepsilon_{r+1\nu}-\varepsilon_{r\nu'}+
\varepsilon_{\ss B}-\hbar\omega-i\lambda }\right\rangle\right\rangle , 
\end{equation}	
where $Q_{r,r'}^{\nu\nu '}=\left|\int d{\bf x}\psi_{r{\bf x}}^{\nu ~*}\psi_{r'
{\bf x}}^{\nu'}\right|^2$ is the overlap factor. Thus, we have evaluated the 
expression for the response with the scattering processes exactly taken into 
account. 

Below we consider the absorption coefficient introduced according to 
$\alpha_{\omega}=(4\pi /c\sqrt{\epsilon})Re\sigma_{\omega}$, where $\epsilon$
is the dielectric permittivity, which is supposed uniform across the structure.
In order to perform averaging in Eq.(9), we use the spectral density function
in the $r$-th QW determined as ${\cal A}_{r\varepsilon}({\bf x},{\bf x'})=
\sum_{\nu}\psi_{r{\bf x'}}^{\nu ~*}\psi_{r{\bf x}}^{\nu }\delta (\varepsilon_{r
\nu}-\varepsilon )$ \cite{11}, so that $\alpha_{\Delta\varepsilon}$ is written 
as follows:
\begin{eqnarray}
\label{10} 
\alpha_{\Delta\varepsilon}=\frac{2(2\pi ev_{\ss\bot})^2}{c\sqrt{\epsilon}
\omega L^3}\int_{-\infty}^{\infty}d\varepsilon (f_{\varepsilon -\Delta
\varepsilon}-f_{\varepsilon}) \\
\times\int d{\bf x}\int d{\bf x'}\sum_r\langle\langle{\cal A}_{r+1\varepsilon}
({\bf x},{\bf x'}){\cal A}_{r\varepsilon -\Delta\varepsilon}({\bf x'},
{\bf x})\rangle\rangle  \nonumber
\end{eqnarray}
with $\hbar\omega\simeq\varepsilon_{\ss B}$ in the resonant approximation.

We turn now to averaging over short-range and large-scale potentials taking
into account that we are considering SL under a homogeneous bias voltage. Due
to this the averaged characteristics of scattering processes, both for 
homogeneous and inhomogeneous mechanisms, do not dependent on the QW number 
$r$. It is convenient to use the Wigner representation and the average of the 
spectral functions in (10) takes the form:
\begin{eqnarray}
\label{11} 
\int\int\frac{d{\bf x}d{\bf x'}}{L^3}\sum_r\langle\langle{\cal A}_{r+1
\varepsilon}({\bf x},{\bf x'}){\cal A}_{r\varepsilon -\Delta\varepsilon}
({\bf x'},{\bf x})\rangle\rangle  \nonumber \\
=\frac{1}{Z}\int\frac{d{\bf p}}{(2\pi\hbar )^2}\langle\langle {\cal A}_{r+1
\varepsilon}({\bf p},{\bf x}){\cal A}_{r\varepsilon -\Delta\varepsilon}({\bf p},
{\bf x})\rangle\rangle .
\end{eqnarray}
Here we took into account that $\langle\langle\ldots\rangle\rangle$ does not 
depend on ${\bf x}$ and $L^{-1}\sum_r=Z^{-1}$. Performing the averaging over 
short-range potential, we write the spectral function $\langle {\cal A}_{r
\varepsilon}({\bf p},{\bf x})\rangle =ImG_{r\varepsilon}^{\ss R}(p,{\bf x})/
\pi$ through the retarded Green's function given by: 
\begin{equation}
\label{12} 
G_{r\varepsilon}^{\ss R}(p,{\bf x}) =(\varepsilon_p-w_{r{\bf x}}-\varepsilon 
-\Sigma )^{-1}.
\end{equation}	
Here $w_{r{\bf x}}$ is a large-scale part of potential in the $r$-th QW and 
$\Sigma$ is the self-energy function arising from the short-range scattering
(see similar calculations in \cite{14}). Below we consider the case of 
scattering by zero-radius centers when $Im\Sigma$ does not depend on 
$\varepsilon$, $p$ or $\bf x$. $Re\Sigma$, which is logarithmically divergent 
without a small-distance cutoff, is included into the detuning energy 
$\Delta\varepsilon$, so that the only homogeneous broadening contribution,
$-i\gamma$, appears in the denominator of Rq. (12). Performing the averaging 
over large-scale potentials we write the spectral density in the integral form:
\begin{equation}
\label{13} 
A_{\varepsilon}(\varepsilon_p)=\int_{\infty}^0\frac{dt}{2\pi\hbar}e^{i
(\varepsilon_p-\varepsilon -i\gamma )t/\hbar}e^{-(\Gamma t/\hbar )^2/2}
+c.c. ,
\end{equation}
where $\Gamma =\sqrt{\langle w_{r{\bf x}}^2 \rangle}$ is the inhomogeneous 
broadening energy. Using the in-plane isotropy of the problem, we finally
transform Eq.(10) into
\begin{eqnarray}
\label{14} 
\alpha_{\Delta\varepsilon}\simeq\frac{e^2}{\hbar c}\frac{4\pi m
{\rm v}_{\ss\bot}^2}{\sqrt{\epsilon}\varepsilon_{\ss B}Z}\int_{-\infty}
^{\infty}d\varepsilon \int_0^{\infty}d\xi \\
\times A_{\varepsilon -\Delta\varepsilon /2}(\xi )A_{\varepsilon +\Delta
\varepsilon /2}(\xi )(f_{\varepsilon -\Delta\varepsilon /2}-f_{\varepsilon 
+\Delta\varepsilon /2}) . \nonumber
\end{eqnarray}
Note, that for the collisionless case the product of spectral functions under 
the integrals is transformed into $\delta (\xi -\varepsilon +\Delta\varepsilon 
/2)\delta (\xi -\varepsilon -\Delta\varepsilon /2)$.

\begin{figure}
\begin{center}
\includegraphics{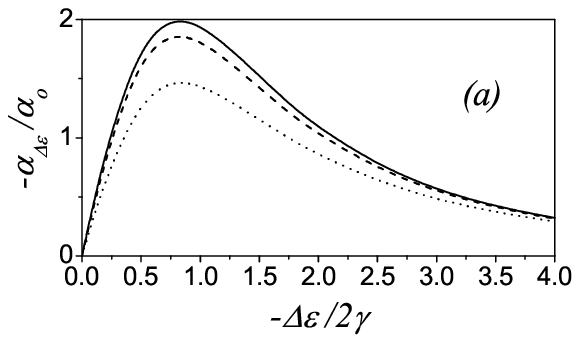}
\includegraphics{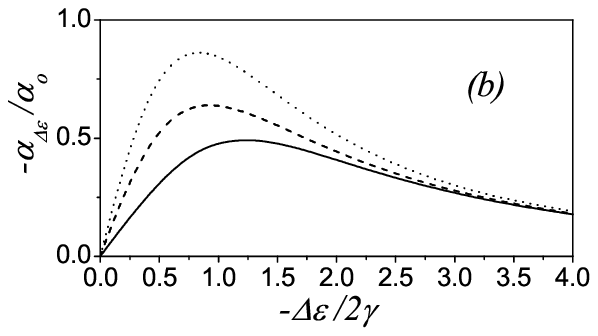}
\end{center}
\addvspace{-1 cm}\caption{Dimensionless gain $-\alpha_{\Delta\varepsilon}
/\alpha_o$ versus detuning energy $-\Delta\varepsilon /\gamma$ for the 
homogeneous broadening case ($\Gamma =0$). Panels ($a$) and ($b$) correspond 
to degenerate ($\mu /\gamma =$3) and non-degenerate ($\mu /\gamma =$ -1) 
electrons. Solid, dashed, and dotted curves correspond to $T_e/\gamma =$0.3, 
1, and 3 respectively.}
\label{fig.2}
\end{figure}

\begin{table}
\begin{tabular}{|c|c|c|c|c|c|c|}
\hline
$T_e/(\gamma ,\Gamma )$ & $\mu /\gamma$=3, $\Gamma$=0 & $\mu /\gamma$=-1, 
$\Gamma$=0 & $\mu /\Gamma$=3, $\gamma$=0 & $\mu /\Gamma$=-1, $\Gamma$=0 & 
$\mu /2\gamma$=3, $\Gamma =\gamma$ & $\mu /2\gamma$=-1, $\Gamma =\gamma$ \\
 \hline 0.3 & 3.3$\cdot 10^{16}$ & 8.8$\cdot 10^{15}$ & 4$\cdot 10^{16}$ & 
1.6$\cdot 10^{15}$ & 4$\cdot 10^{16}$ & 8.2$\cdot 10^{15}$ \\ \hline
1 & 3.4$\cdot 10^{16}$ & 1.1$\cdot 10^{16}$ & 4.1$\cdot 10^{16}$ & 
5.4$\cdot 10^{15}$ & 4.1$\cdot 10^{16}$ & 1.3$\cdot 10^{16}$ \\ \hline
3 & 4.2$\cdot 10^{16}$ & 1.5$\cdot 10^{16}$ & 5.2$\cdot 10^{16}$ & 
2.2$\cdot 10^{16}$ & 5$\cdot 10^{16}$ & 3.4$\cdot 10^{16}$ \\ \hline
\end{tabular}
\caption{\label{tab:table1}3D concentrations, measured in cm$^{-3}$, versus 
dimensionless $\mu$ and $T_e$ for the cases plotted in Figs.2 ($\Gamma =0$), 
3 ($\gamma =0$), and 4 ($\Gamma =\gamma$).}
\end{table}

Further, we calculate the spectral dependencies $\alpha_{\Delta\varepsilon}/
\alpha_o$ given by Eq.(14) with the use of the quasi-equilibrium Fermi 
distribution $f_{\varepsilon}$. The characteristic absorption $\alpha_o$ is 
introduced here according to $\alpha_o=(e^2/\hbar c)m{\rm v}_{\ss\bot}^2/
\sqrt{\epsilon}\varepsilon_{\ss B}Z$ and $\alpha_{\Delta\varepsilon}=-
\alpha_{-\Delta\varepsilon}$, so that we consider only the region $\Delta
\varepsilon <0$. First, we examine the cases of homogenous ($\Gamma =0$) and
inhomogeneous ($\gamma =0$) broadening. The dimensionless gain is plotted for 
the cases of degenerate [Figs.2$(a)$ and 3$(a)$] and nondegenerate [Figs.2$(b)$ 
and 3$(b)$] electrons with the concentrations given in Table I. According to 
these data, concentration increases with $T_e$ for all cases but the peak 
gain varies in a different manner: quenching or enhancement of gain occures
for degenerate or non-degenerate electrons respectively. High-energy tails of
gain and lower peak values take place for the homogeneous broadening case, when
(13) has a Lorentzian shape. For the inhomogeneous broadening case, 
$A_{\varepsilon}(\xi )$ has a Gaussian shape and $\alpha_{\Delta\varepsilon}$ 
appears to be a sharper function. In Fig.4 we present the case $\gamma =
\Gamma$, plotting the dimensionless gain versus $\Delta\varepsilon /2(\gamma +
\Gamma )$, where $2(\gamma +\Gamma )$ is the total width of (13). One can see 
both the same style of spectral dependencies and the same 
temperature/concentration dependencies.

Next we turn to estimates of the maximal gain for $GaAs/Al_{0.3}Ga_{0.7}
As$-based BSL with $T=$0.5 meV, corresponding to the barrier width of 6 nm and
$Z=$15 nm. For the level splitting energy $\varepsilon_{\ss B}=$ 10 meV, 
which is correspondent to the transverse field $F=$6.7 kV/cm, one obtains
$\alpha_o\simeq$6.6 cm$^{-1}$, so that the peak gain appears to be between
5 and 20 cm$^{-1}$ for different parameters used in Figs.2-4. Note, that 
$\alpha_o\propto T^2$ in the framework of the tight-binding approximation and 
gain increases rapidly for the narrow barrier case, for example gain exceeds 
the experimental data \cite{3}, if $T=$1 meV.

\begin{figure}
\begin{center}
\includegraphics{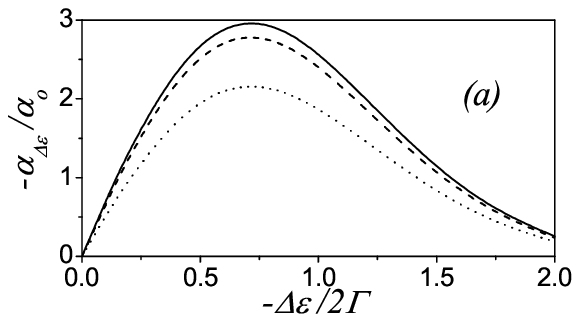}
\includegraphics{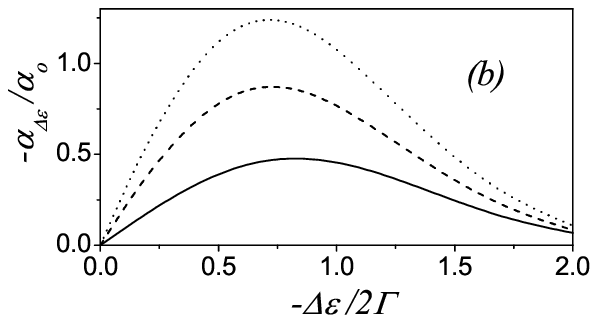}
\end{center}
\addvspace{-1 cm}\caption{The same as in Fig.2 for the inhomogeneous broadening
case ($\gamma =0$) depending on parameters $\mu /\Gamma$ and $T_e/\Gamma$.}
\label{fig.3}
\end{figure}

\begin{figure}
\begin{center}
\includegraphics{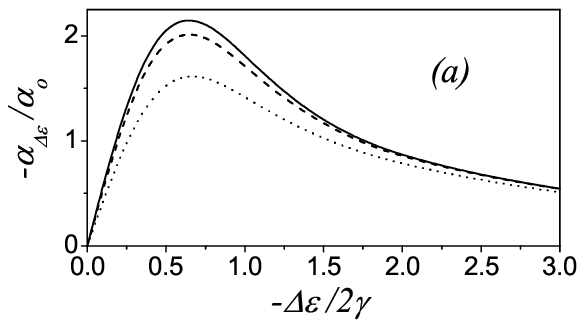}
\includegraphics{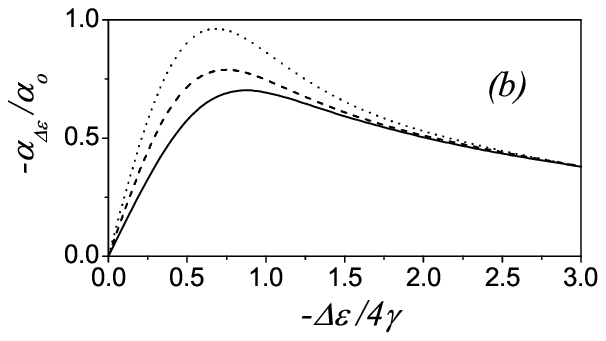}
\end{center}
\addvspace{-1 cm}\caption{Dimensionless gain $-\alpha_{\Delta\varepsilon}
/\alpha_o$ versus $-\Delta\varepsilon /2(\gamma +\Gamma )$ for the case $\gamma 
=\Gamma$. Panels ($a$) and ($b$) correspond to degenerate ($\mu /2\gamma =$3) 
and non-degenerate ($\mu /2\gamma =$ -1) electrons. Solid, dashed, and dotted 
curves correspond to $T_e/2\gamma =$0.3, 1, and 3 respectively.}
\label{fig.4}
\end{figure}

Let us discuss the main assumptions used. The tight-binding approach is valid 
under the condition $\varepsilon_{\ss B}\gg 2T$ which is satisfied for the 
numerical estimates performed; note, that beyond the Born approximation the 
broadening can be comparable with the electron energy determined through $\mu$ 
and $T_e$. We restrict ourselves to the case of homogeneous field and 
concentration distributions neglecting a possible domain formation due to the
negative differential conductivity at low frequencies \cite{15}. One can avoid
instabilities in a short enough BSL because the THz modes propagate in the
in-plane directions. In spite of the general expressions (10-12) are written 
through an arbitrary self-energy function $\Sigma$, the final calculations 
were performed for the model included scattering by zero-radius centers and 
large-scale potential. Such a model describes the interplay between 
homogeneous and inhomogeneous broadening with the use of statistically 
independent random potentials in each QW. The Coulomb correlations, which 
modify the response as the concentration increases, are not taken into account 
here. This contribution, as well as consideration of intermediate-scale 
potential, require a special consideration in analogy with the case of a
single QW \cite{16}.

In conclusion, we have considered the resonant intersubband response of a BSL
and have described gain without inversion beyond the Born approximation. 
It seems likely that this contribution can be found experimentally and more
detailed numerical calculations are necessary in order to estimate a
potential for applications. 

\textbf{Acknowledgment}: {This work was supported by Science Foundation 
Ireland. \\
$^{*}$ E-mail: {\rm ftvasko@yahoo.com}.\\
On leave from: Institute of Semiconductor Physics, Kiev, NAS
of Ukraine, 252650, Ukraine
.}


\begin{thebibliography}{16}
\bibitem{12}
It should be noted that the exact description of scattering prosesses 
distinguishes the calculations performed from Ref.9; moreover, the spectral 
dependencies given by Eqs.(13,14) for the homogeneous broadening case do not 
coincide with Eqs.(19,24) in \cite{9}.
\end{thebibliography}
\end{document}